# Timing Matters: How Using LLMs at Different Timings Influences Writers' Perceptions and Ideation Outcomes in AI-Assisted Ideation

Peinuan Qin
School of Computing, National University of Singapore
Singapore, Singapore
e1322754@u.nus.edu

Chi-Lan Yang
Graduate School of Interdisciplinary Information Studies, The University of Tokyo
Tokyo, Japan

Jingshu Li
Computer Science, School of Computing, National University of Singapore
Singapore, Singapore

Jing Wen
National University of Singapore
Singapore, Singapore

Yi-Chieh Lee
Computer Science, National University of Singapore
Singapore, Singapore

## Abstract

Large Language Models (LLMs) have been widely used to support ideation in the writing process. However, whether generating ideas with the help of LLMs leads to idea fixation or idea expansion is unclear. This study examines how different timings of LLM usage - either at the beginning or after independent ideation - affect people's perceptions and ideation outcomes in a writing task. In a controlled experiment with 60 participants, we found that using LLMs from the beginning reduced the number of original ideas and lowered creative self-efficacy and self-credit, mediated by changes in autonomy and ownership. We discuss the challenges and opportunities associated with using LLMs to assist in idea generation. We propose delaying the use of LLMs to support ideation while considering users' self-efficacy, autonomy, and ownership of the ideation outcomes.

## 1 Introduction

The integration of Large Language Models (LLMs) into the ideation process has sparked debates about whether their use fosters idea expansion or leads to idea fixation. While LLMs have demonstrated the ability to enhance creativity by expanding users' conceptual space and increasing creative self-efficacy [9, 20, 34, 39], they may also undermine critical thinking and originality by inducing cognitive fixation on AI-generated content [19, 50, 62].

Empirical evidence has established that the timing of providing creative examples significantly affects the creation outcomes [30, 31, 41, 48, 49, 51]. Providing examples at a suitable timing is beneficial for promoting creative outcomes and alleviating idea fixation [31], while doing so at an inappropriate timing exacerbates the rigidity of creativity [41]. For example, Kohn and Smith demonstrated that being exposed to others' ideas during brainstorming reduces idea diversity and leads to idea fixation [30]. However, as LLMs serve as powerful example generators, how different timings of their usage influence idea generation has not been thoroughly examined.

At the same time, due to the powerful capabilities of LLM, their integration into creative tasks often changes human autonomy while completing the tasks [2, 25, 32, 37]. Research [12, 65] suggests that when individuals engage in tasks out of intrinsic interest or desire, rather than external pressure, they demonstrate greater effort, persistence, and innovation. Furthermore, autonomy plays a critical role in perceptions associated with creative processes, such as people's ownership and credit attribution of the creative outcome [18, 52, 55, 65]. Thus, examining the timing of LLM usage is crucial for understanding how it influences users' ideation outcomes and perceptions in AI-assisted ideation.

This study examined how using LLM at different timings during the ideation process impacts individuals' creative self-efficacy, autonomy, ownership, credit attribution, and ideation outcomes. Specifically, we address the following research questions:

RQ1: How does ideating with LLM at different timings (*before* or *after* independent ideation) influence ideation outcomes (e.g., quantity of supporting ideas, similarity with LLM ideas, and personal idea diversity)?

RQ2: How does ideating with LLM at different timings (*before* or *after* independent ideation) impact participants' perceptions of their creative self-efficacy, autonomy, ownership, and credit attribution?

RQ3: What is the mediation mechanism, and how does it transmit the effect of LLM usage timing to ideation outcomes and credit attribution?

We conducted a controlled experiment involving 60 participants who were tasked with generating at least three supporting ideas on health-related topics. Results revealed that introducing LLM *after* participants had engaged in independent ideation increased the quantity of ideas and decreased the similarity with LLM ideas. Conversely, using LLM at the beginning led to idea fixation, reduced creative self-efficacy, and diminished the level of credit participants attributed to themselves. Participants who used LLM later exhibited greater autonomy and ownership, which fully mediated the improvement in ideation outcomes.

Our work provides the following contributions.

(1) An empirical understanding of how the timing of LLM usage influences individuals' perceptions and ideation outcomes through a mixed-method approach.
(2) Insights into the role of perceived autonomy in mediating ideation outcomes and design implications for fostering user autonomy in AI-assisted ideation.

## 2 Related Work

### 2.1 Using AI to Assist Creative Tasks

With the development of LLMs, AI has been used to support various types of creative tasks, including writing [9, 26, 39, 54, 68], drawing [44, 62, 69], or making songs with AI [24]. Despite the prevalence of AI-assisted creative work, there are ongoing debates about whether working with AI to complete a creative task leads to *idea expansion*—a process where AI inspires novel ideas and enhances creative outcomes—or *idea fixation* [1, 27, 61, 67], where users focus too narrowly on AI-generated suggestions, narrowing divergent thinking and constraining individual creativity.

On the one hand, studies have shown that AI enhances human creativity [7, 8, 14, 15, 28, 38]. For instance, research demonstrates that the ability of LLMs to generate cross-domain analogies assists individuals in breaking away from fixed thinking patterns, reframing problems, and igniting innovative solutions [14]. Furthermore, studies have demonstrated that AI can enhance people's creative self-efficacy [38] and has been seamlessly integrated into various content creation processes, such as brainstorming [57], collaborative slogan writing [9], and story writing [54, 68]. For example, a recent study showed that co-writing with AI enhanced people's creative self-efficacy and improved creative outcomes in a poetry-writing task [38]. Jaussi et al. also discussed that AI reduced barriers to performing creative tasks along with improving creative self-efficacy [28]. These findings suggested that AI has the potential to inspire users, leading to idea expansion [33, 47, 53].

On the other hand, some other research suggested that AI may hinder the creative process by diminishing creative process, independent thought, and originality [13, 19, 50, 62]. For instance, Dergaa et al. [13] introduced the concept of "AI-induced cognitive atrophy", suggesting that over-reliance on AI diminishes mental engagement and independent thinking. This could lead to a decline in people's cognitive functions, including creativity. Anderson et al. proposed the "homogenization" issue of LLM-assisted creative support, where they revealed that users who employed LLMs during the group ideation process generated less semantically diverse ideas collectively than those who used non-LLM-based supporting tools [1]. Additionally, LLM users reported feeling less responsible for the ideas they generated, attributing the ideas more to the LLM-based creative supporting tools than to themselves [1]. Neuroscience research also provided consistent findings, suggesting that reduced cognitive exertion weakens neuronal connections, thus stifling human-driven innovation [19]. Furthermore, Wadinambiarachchi et al. provided empirical evidence demonstrating that AI assistance during the ideation phase led to idea fixation, with participants generating fewer and less diverse ideas compared to those who worked without AI. This line of literature suggests that AI can also lead to idea fixation.

Taken together, these studies highlighted the inconclusive findings for the use of AI in creative tasks. To address this gap, we explore how different timings of using LLM in an ideation process impact the creative process, aiming to optimize the support of LLM in the ideation process.

### 2.2 Timing of Introducing Examples in Creative Tasks

Related work has demonstrated that the timing of example provision significantly affects people's creative process [30, 31, 41, 48, 49, 51]. According to the Search for Ideas in Associative Memory (SIAM) theory, examples can have both positive and negative impacts on creative outcomes, depending on when they were introduced [41, 51]. For instance, research suggested that early or repeated exposure to examples improved creativity [31]. Conversely, certain educational studies [48, 49] found that students benefited more from learning resources when they first comprehend concepts on their own before being introduced to such materials. Nijstad and Stroebe stressed that providing examples at the wrong time can disrupt the cognitive flow, resulting in idea loss and lower creativity [41]. For example, Kohn and Smith demonstrated that exposure to others' ideas during brainstorming reduces idea diversity and leads to fixation [30]. Since LLMs often function as example generators, it is important to investigate how different timings of LLM usage influence idea generation.

Additionally, some evidence shows that an ongoing shift in human autonomy [2, 25, 32, 37, 64] because of the increasingly growing ability of AI. People gradually treat AI as a coordinator rather than as a mere auxiliary tool when using AI in completing a creativity task [25, 37]. Therefore, the way AI intervenes in the creative process, such as timing, may alter people's sense of autonomy. While engaging in creative activities, autonomy is crucial as it encourages individuals to take ownership of their ideas [18, 55] and feel responsible for their creative decisions [10, 52], while also increasing their creative self-efficacy [36, 65]. Studies have shown that when people perceive greater autonomy in their tasks, they engage more deeply, explore novel ideas, and produce higher-quality creative outcomes [36, 65]. Furthermore, McGuire et al. have found that when participants co-write with AI rather than simply editing

AI-generated content, they experience higher levels of creative self-efficacy and produce more original work [38], and speculate that this result is related to the sense of autonomy in discussion.

Therefore, this literature motivated us to investigate how using LLM at different timings influences people's autonomy, ownership, and their tendency to assign credit to themselves or LLM for creative outcomes.

## 3 Method

### 3.1 Study Design

The study aims to investigate (1) how using LLM at different timings influences ideation outcomes (e.g., the quantity of ideas and similarities with LLM), (2) how using LLM at different timings influences people's perceptions (e.g., autonomy, ownership, and creative self-efficacy, credit attribution) in the writing process, and (3) how the effect of LLM usage timing transmits to changes in ideation outcomes and credit attribution through mediation factors.

A between-subjects design was used to explore these questions. We compared $C_{before}$ and $C_{after}$, which refer to the LLM joining the ideation process *before* and *after* participants independently generated ideas, respectively. In an ideation task, we asked participants to generate supporting ideas for their claims when writing health-related topics. Participants were not required to have specific expertise in these areas. We instructed them to generate supporting ideas based on their current understanding of the domain and to use logical reasoning without worrying about whether their ideas were factual. The purpose of the task was not to write a factual article. To ensure the generalizability and robustness of our findings, two specific topics were selected for the writing task. One topic is about *"Excessive daytime sleepiness increases the risk of developing certain health conditions"* and the other is about *"Applying urine to a jellyfish sting not only fails to ease the pain but can even worsen the sting"*. These health-related topics were chosen due to their prevalence among our targeted participants. We asked participants to ideate as many supporting ideas as they could to substantiate the claim for the given writing topic.

### 3.2 Participants

We recruited a diverse group of 60 participants through Prolific[1]. These individuals were randomly assigned into two conditions, with 30 participants per condition. The demographic makeup of the sample was internationally diverse, covering an age range of 21 to 50 years (M = 31.167, SD = 8.061). Among all participants, 26 (43.3%) identified themselves as female and 34 (56.7%) as male. Regarding educational level, the majority of participants (67%) indicated having obtained at least a bachelor's degree. Each participant received a compensation of 5.85 USD for participating in an average of a 30-minute task. The payment was decided based on the platform we used and was considered a fair price, according to Prolific. The study was approved by the Institutional Review Board (IRB) of the authors' institute.

### 3.3 Procedure

Before the study began, the participants were instructed about the ideation tasks, which required them to generate supporting ideas for given topics and sign the consent forms. Next, they completed the pre-survey, which asked for their demographic information. After that, participants practiced writing supporting ideas with our prototype for 5 minutes. They were instructed about how to work with LLM using our prototype for idea generation and provided with some effective tips for prompting the LLM in the prototype. The goal of this practice session was to bring participants' familiarity with prompting using the prototype to a similar level, and we ensured everyone received an equal amount of training prior to the main ideation task. In this practice session, we asked participants to generate as many distinct and relevant supporting ideas as possible for the topic: *"Think of the benefits of using social media."*

Afterward, the participants were randomly assigned to two conditions ($C_{before}$ or $C_{after}$) to complete the writing task, writing supporting ideas for one of the writing topics. The topic assignment was counterbalanced. In $C_{before}$, participants used LLM from the very beginning to generate supporting ideas for the given topic, without putting effort into independent ideation. In $C_{after}$, participants first generated at least three distinct and original supporting ideas independently before being allowed to use LLM to supplement or expand their ideas. In total, the ideation task lasted approximately twenty minutes.

After the ideation task, participants ended the experiment by filling out a post-survey to share their perception of the ideation process. This process lasted about 5 minutes. The whole online experiment took about 30 minutes in total, and the procedure can be found in Figure 1.

### 3.4 Prototype for Generating Ideas with LLM

We developed a web application for participants to generate supporting ideas with LLM at different stages of the ideation process. We used Vue.js[2] with Javascript[3] for front-end development, and took Django[4] with Python[5] to build the back-end server. This application was installed on Heroku[6] for public access during data collection. Additionally, we adopted OpenAI's GPT-4[7] API to build the streaming content generation feature, aiming to maintain smooth interaction and high-quality LLM responses.

*3.4.1 Interface.* Figure 2 illustrates the interface participants used in the study. The interface was divided into three main components: (A) *Instruction Component*, which displayed the tasks to be completed at each stage. The instructions were dynamically updated as the task progressed. (B) *Ideation Component*, where participants generated, modified, and submitted their ideas. By default, the interface provided three input fields. Participants can click the 'Add New' button to add more input fields or use the 'Delete' button next to each idea to remove it. Clicking 'Reset All' clears all ideas and restores the interface to its initial state. (C) *AI Generation Component*, whose visibility depended on the participant's assigned experimental condition. For participants in the $C_{before}$ condition, the AI generation component remained visible throughout the main task. For participants in the $C_{after}$ condition, the 'Display AI' button was initially disabled. It was enabled once participants had created at least three ideas and clicked 'Save Idea'. Once enabled, participants can click the button to reveal the full AI Generation Component.

---

[1] https://www.prolific.co/
[2] https://vuejs.org/
[3] https://en.wikipedia.org/wiki/JavaScript
[4] https://www.djangoproject.com/
[5] https://www.python.org/
[6] https://www.heroku.com/
[7] https://openai.com/index/gpt-4/

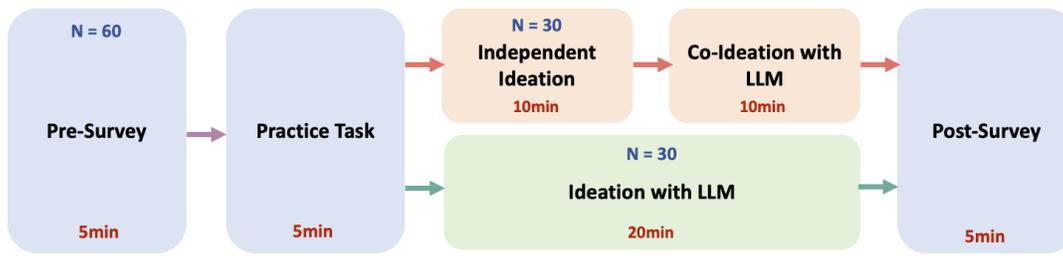

Figure 1: Experiment workflow of our study. Participants experienced a pre-survey, a practice task, an ideation task, and a post-survey during the study. The timing for participants to use LLM during the ideation task differed across the two experimental conditions. In the condition colored in orange, LLM was introduced after independent ideation, while in the condition colored in green, LLM was available at the beginning of the ideation task.

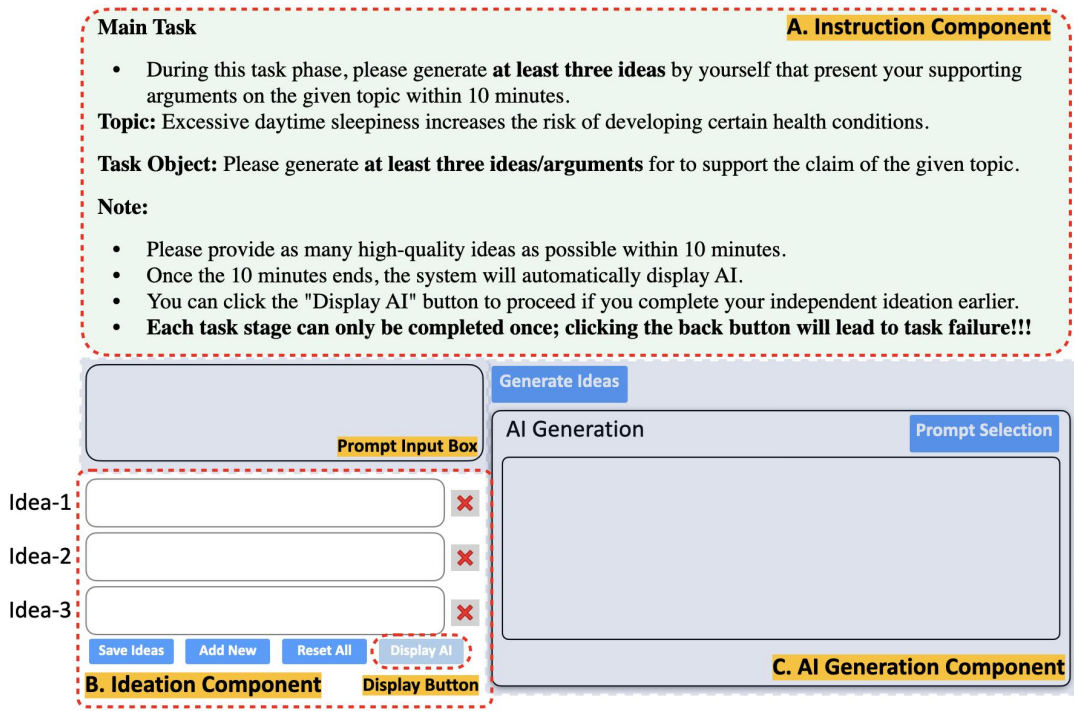

Figure 2: Overview of the task interface presented to participants during the ideation task. The interface consists of three main components: (A) Instruction Component, providing dynamic task guidance and requirements based on the current task stage; (B) Ideation Component, where participants create, edit, and manage their ideas using input fields, and interact with buttons for saving, adding, resetting, or deleting ideas; and (C) AI Generation Component, enabling participants to use AI-assisted ideation tools that include prompt input and AI-generated suggestions display panel, which are visible based on the experimental condition.

The AI Generation Component included a 'Prompt Input Box', located above the ideation component. Participants can click the 'Prompt Selection' button to choose from pre-designed prompt templates described in Section 3.4.2. They can modify the selected template or create a personalized prompt directly. After finalizing the prompt, participants can click the 'Generate Ideas' button to receive AI-generated ideation results, which will be displayed in the right-hand area for reference.

Table 4 (see Appendix A) presents the outcome for a participant from condition $C_{before}$ interacting with the prototype for idea generation. Throughout this process, AI provided nine ideas as references, and the participant concluded by submitting three ideas.

*3.4.2 Prompt Design for AI-assisted Ideation.* Participants accessed the LLM to generate supporting ideas for the ideation task at different stages. To enable participants to easily generate supporting ideas with LLM, we predefined three default prompts as examples for participants to refer to. To standardize the responses and ensure consistency across participants, each prompt was designed to elicit four concise supporting ideas (each within 50 words). This format minimized cognitive overload and allowed participants to focus on evaluating and refining ideas. The following showed three example prompts designed to guide participants in generating supporting ideas for one of the topics: *"Excessive daytime sleepiness increases the risk of developing certain health conditions"*.

- **Default Prompt 1:** *Please give me four supporting ideas on the topic "Excessive daytime sleepiness increases the risk of developing certain health conditions." Each within 50 words.*
- **Default Prompt 2** *Please elaborate on the long-term excessive daytime sleepiness on an individual's well-being by listing four ideas/arguments, each within 50 words.*
- **Default Prompt 3:** *Please provide more convincing insights on the connection between excessive daytime sleepiness and mental health disorders by listing four ideas, each within 50 words.*

We also allowed participants to adjust the predefined prompts or create their own prompts to generate supporting ideas for the given writing topic.

## 3.5 Measurement

To answer our research questions, we collected quantitative data from system logs and the questionnaire, as well as qualitative data from participants' open-ended responses.

*3.5.1 Task Completion Time.* To check whether participants engaged in the ideation task, we recorded the time from when they began the task until they submitted their ideas. The descriptive data showed that participants in $C_{before}$ completed their tasks in an average time of 575.033 seconds (SD = 338.897), while those in $C_{after}$ took an average of 711.433 seconds (SD = 330.966).

*3.5.2 Quantity of Supporting Ideas (RQ1).* We defined a "supporting idea" as any distinct idea a participant had generated to substantiate the claim for the given writing topic. To calculate the total quantity of supporting ideas, we counted each distinct idea provided by the participant, regardless of whether it was generated independently or with the assistance of the LLM. Redundant or repetitive ideas were not included in the final count.

*3.5.3 Similarity between Ideas from Participant and LLM (RQ1).* We measured the similarity between ideas generated by participants and those generated by the LLM to assess the extent of LLM's influence on ideation. Higher similarity scores indicated greater overlap with LLM-generated ideas, suggesting a greater reliance on LLM, while lower scores reflected more original and independent ideas. We used the term frequency-inverse document frequency (TF-IDF)[8] combined with cosine similarity to measure the overlap between participant-generated and LLM-generated ideas since the TF-IDF directly captured term overlap.

The similarity calculation involved the following steps. The details can be found in Table 1. *(1) Text preprocessing:* Both participant-generated ideas and LLM-generated ideas were preprocessed by removing stop words, stemming or lemmatizing words, and converting text to lowercase to ensure consistency in comparison. *(2) TF-IDF vectorization:* The preprocessed text was converted into TF-IDF vectors, which captured the importance of terms in the context of the entire set of ideas generated during the experiment. *(3) Similarity calculation:* For each participant-generated idea, the cosine similarity score was calculated with each idea generated by LLM, all of which comprised the individual similarity matrix. Each column in the matrix was treated as a vector, with its magnitude representing the degree of overlap between a single writing idea and all other LLM suggestions. Averaging all magnitudes of column vectors can finally calculate the score representing the participant's reliance on the LLM.

*3.5.4 Participant's Idea Diversity.* In creativity research, idea diversity is commonly used to evaluate the breadth and originality of ideation [11, 23]. Building on [23], we assessed diversity using the semantic differences between each idea generated by each individual.

Given one's idea set, we called the OpenAI embedding API[9] using text-embedding-ada-002 to embed each idea, followed by computing a cosine similarity matrix, where each element represents the similarity between a pair of ideas. To focus on unique pairs and exclude self-similarity, we extract the upper triangular portion of the matrix (excluding the diagonal). The 'average similarity' across these pairs is computed, and the diversity score is defined as 1 - 'average similarity'. This approach ensures that higher semantic differences among ideas correspond to higher diversity scores.

*3.5.5 Creative self-efficacy (RQ2).* Creative self-efficacy refers to an individual's belief in their ability to produce creative outcomes or generate novel and useful ideas during a task [4, 59]. It is a specific form of self-efficacy [5], which is the broader concept of confidence in one's ability to succeed in a particular domain. We measured creative self-efficacy using the five items adapted from [59], containing *"I have confidence that I solved the task creatively", "I had a knack for further developing the ideas of others", "I feel that I was good at generating novel ideas during the task", "I had a lot of good ideas during the task",* and *"I had good imagination during the task"*. Participants answered the items with a 7-point Likert scale, where one indicates strongly disagree and seven indicates strongly agree. We calculated the score for creative self-efficacy by averaging the five items. The Cronbach's alpha for the five items is $\alpha = 0.900$.

---

[8] We also tried transformer-based embeddings (using the OpenAI text-embedding-ada-002 model to embed each idea) for semantic similarity calculation while keeping the other steps the same. It demonstrated a similar trend in the statistical analysis across conditions, which did not alter the conclusion reached using TF-IDF. Please refer to the Appendix A to see the result of using transformer-based methods to calculate similarity scores (see Figure 7).
[9] https://platform.openai.com/docs/api-reference/embeddings

Table 1: Pseudocode for calculating similarity scores between participant-generated and LLM-generated ideas.

| Pseudocode Description |
| --- |
| **Input**: Participants $\mathbb{P}$, LLM-generated ideas $\mathbb{L}_i$ for each $P_i$, participant-generated ideas $\mathbb{I}_i$ for each $P_i$ |
| **Output**: Similarity score for each $P_i$ |
| 1: Initialize an empty vocabulary $\mathbb{V}$ |
| 2: For each participant $P_i \in \mathbb{P}$ |
|    2.1: Collect words from both $\mathbb{L}_i$ and $\mathbb{I}_i$ into $\mathbb{V}$ to construct the lexicon |
| 3: Construct a TF-IDF vectorizer using $\mathbb{V}$ |
| 4: For each participant $P_i \in \mathbb{P}$ |
|    4.1: Initialize an empty similarity matrix $S_i$ of size $j \times k$ |
|    4.2: For each participant-generated idea $I_{ij} \in \mathbb{I}_i$ |
|       4.2.1: Compute cosine similarity between $I_{ij}$ and each LLM-generated idea $L_{ik} \in \mathbb{L}_i$ |
|       4.2.2: Store results in similarity matrix $S_i$ |
| 5: Initialize an empty list $\mathbb{D}_i$ to store individual idea similarity scores |
| 6: For each participant $P_i \in \mathbb{P}$ |
|    6.1: For each idea $I_{ij} \in \mathbb{I}_i$ |
|       6.1.1: Compute the magnitude of the TF-IDF vector $|\vec{v}_j|$ as the similarity score $D_{ij}$ |
|       6.1.2: Append $D_{ij}$ to $\mathbb{D}_i$ |
|    6.2: Compute the average of $\mathbb{D}_i$ to obtain the overall similarity score for $P_i$ |
| 7: Return similarity scores for all participants |

*3.5.6 Sense of Ownership in Ideation (RQ2).* The extent to which participants felt like they were the authors of the ideation outcome indicates their sense of ownership. We assess participants' sense of ownership with the following five items on a 7-point Likert scale (1 indicates strongly disagree and 7 indicates strongly agree): *"I feel like I am the author of the text;" "I am the main contributor to the content of the resulting text;" "I have made substantial contributions to the content of the resulting text;" "I am accountable for all aspects of the text;" "I am responsible for at least part of the text."* We calculated the score for the sense of ownership by averaging the five items. The Cronbach's alpha for the five items is $\alpha = 0.867$.

*3.5.7 Sense of Autonomy in Ideation (RQ2).* Autonomy refers to the degree to which individuals feel they have control over their actions and decisions, reflecting their ability to act according to their own will without external constraints. In the context of ideation with LLM assistance, it denotes how much control participants perceived over their ideation outcome in the writing process. We adapted and modified the scales in [16] and asked the following two questions: *"I felt like I was writing the text and the AI was assisting me;"* and *"I felt like AI was writing the text and I was assisting."* We calculated the score for the sense of autonomy by averaging the two items. The Cronbach's alpha for the two items is $\alpha = 0.732$.

*3.5.8 Credit attribution (RQ2).* In our study, credit attribution refers to participants' perceptions of their own and the LLM's contributions to generating supporting ideas during the ideation task. This was measured by asking two questions to participants with a 7-point Likert scale: *(1) self-credit, "How much credit do you think you have in writing the final text?"* and the other for *(2) credit to LLM, "How much credit do you think AI has in writing the final text?"* A score of one indicated very little credit, while a score of seven indicated very much credit.

*3.5.9 Open-ended Questions.* The open-ended questions were used to triangulate the quantitative results, and we focused on understanding (1) how generating ideas with LLM affected participants' ideation processes and perceptions, as well as (2) how participants interpreted and assigned credit to themselves and LLM when generating ideas with LLM at different timings.

For the data analysis, we employed a thematic analysis method [6]. This began with a detailed familiarization stage, where we carefully reviewed the participants' responses to understand their interactions with LLM. The subsequent phase involved creating initial codes by identifying and labeling key patterns and insights. We identified participants' positive and negative remarks about collaborating with the LLM to capture the general trend of how LLM influenced their thinking and ideation process. A positive remark was defined as any acknowledgment of the LLM's positive influence when using LLM to generate ideas, while a negative remark involved any concerns or doubts expressed about the use of LLM. We also identified codes that participants discussed, such as autonomy, ownership, and credit attribution, when using LLM for generating supporting ideas. These codes were then compared, merged, and refined, resulting in the formation of themes that effectively represented participants' perceptions about generating ideas with LLM.

## 4 Results

This section presents the results of our investigation into the impact of the different timings of LLM usage on ideation outcomes (RQ1), participants' perceptions (RQ2), and mediation effects (RQ3). In the following section, $C_{before}$ refers to the condition where participants began generating ideas immediately with the assistance of LLM, while $C_{after}$ refers to the condition where participants independently generated ideas before using LLM. We used JASP

v0.18.3[10] to analyze the data. For all data based on Likert scale responses, we used non-parametric methods (Mann-Whitney U test) for analysis, as these responses are ordinal in nature and do not necessarily meet the assumptions of parametric tests. For other log-based data, we used the independent Student's t-test, before which we assessed the assumptions of normality and homogeneity of variances using the Shapiro-Wilk test and Levene's test, respectively. The Student's t-test was used when these assumptions were met, and Welch's t-test was applied when violations were detected.

## 4.1 The Impact of LLM Usage Timing on Ideation Outcomes (RQ1)

The result of the t-test showed that the timing of LLM usage had a marginal effect on the number of supporting ideas participants generated in the study ($t(58) = -1.897, p = 0.063, d = -0.490$), and the effect size was moderate ($d = -0.490$), suggesting that the timing of LLM usage may still have practical implications, shown in Figure 3(a). Generating ideas with LLM from the start ($C_{before}$) led participants to write slightly fewer supporting ideas ($M = 3.867; SD = 2.285$) compared to the $C_{after}$ condition ($M = 5.567; SD = 4.345$).

Next, the Welch t-test revealed that the timing of LLM usage had a significant impact on the similarity between participant-generated and LLM-generated ideas ($t(48.36) = 4.778, p = 0.006, d = 0.747$). Specifically, participants in the $C_{before}$ condition had a higher similarity score ($M = 0.899; SD = 0.404$) than in the $C_{after}$ condition ($M = 0.648; SD = 0.250$), which suggested that using LLM from the start made participants have more overlapped ideas with LLM, compared with using LLM after independent ideation (see Figure 3(b)). This suggested that introducing LLM after independent ideation fostered original contributions.

Regarding the participant's idea diversity (see Figure 3(c)), the Welch t-test showed that there was no significant difference between different timings of LLM usage ($t(36.472) = 1.337, p = 0.189, d = 0.345$).

## 4.2 The Impact of LLM Usage Timing on Participants' Perception (RQ2)

To address RQ2, we tested the effect of LLM usage timing on (a) creative self-efficacy, (b) autonomy, (c) ownership, and credit attribution (d and e) with the Mann-Whitney U test. The results are shown in Figure 5.

*4.2.1 Creative Self-Efficacy.* Regarding creative self-efficacy (see Figure 5(a)), the result of the Mann-Whitney U test showed that generating supporting ideas with LLM before independent ideation $C_{before}$ made people perceive significantly lower ($U = 289.500, p = 0.018, r = -0.357$) creative self-efficacy ($M = 4.627; SD = 1.017$), compared to those in the $C_{after}$ condition ($M = 5.300; SD = 0.921$).

Furthermore, Wilcoxon signed-rank tests were employed to assess the change in creative self-efficacy within each condition by comparing their self-efficacy before and after the ideation task (see Figure 4). The analysis revealed that individuals in $C_{before}$ exhibited a significant reduction in creative self-efficacy ($W = 229.5, z = 2.271, p = 0.023, r = 0.530$) from their initial state during the ideation process. Prior to the ideation task their creative self-efficacy statistics were $M = 5.093$ ($SD = 0.985$), which decreased to $M = 4.627$ ($SD = 1.017$) upon task completion (see Figure 4(a)). Nevertheless, for participants in the $C_{after}$ condition, the data suggested that allowing spaces for independent ideation before using LLM did not significantly impact their creative self-efficacy ($W = 215.0, z = 0.273, p = 0.784, r = 0.059$). The creative self-efficacy after ideation ($M = 5.300; SD = 0.921$) showed only a minimal difference from the initial writing ($M = 5.360; SD = 0.898$), shown in Figure 4(b).

These results indicated that using LLMs for ideation from the beginning could negatively impact people's creative self-efficacy. However, allowing spaces for independent ideation before introducing LLMs maintains people's creative self-efficacy.

*4.2.2 Participants' Interpretation of LLM Outputs and Different Levels of Self-Efficacy in Ideation Process.* Qualitative analysis of participants' open-ended comments after the experiment showed that participants thought that writing with LLM both reduced and increased their self-efficacy depending on different interpretations of the outputs generated by LLM.

Participants attributed their lack of confidence and laziness in the ideation process to the powerful ability and convenience of LLM (n = 15 from both conditions). For example, LLM's convenience made some participants feel that their own input was unnecessary, reducing their self-efficacy in engaging with the writing tasks. Participant-37 ($C_{before}$) described this sentiment, stating, *"The AI generates ideas so quickly that it devalues the time that goes into my own thought process and makes me feel like there is no point in putting in the thought, time, or work."* Similarly, Participant-49 ($C_{before}$) reflected on how the LLM serves as a shortcut, possibly allowing them to bypass deeper thinking: *"It has made my idea generation much quicker but possibly made me lazy in terms of coming up with ideas and completing them myself without assistance".*

However, some others (n = 13 from both conditions) mentioned that LLMs increased their self-efficacy in generating ideas, expanding on existing ones, or broadening insights that they had never considered before. One participant mentioned, *"AI helps with generating ideas and editing them efficiently and coherently"* (Participant-58, $C_{before}$), emphasizing the efficiency LLM brought to the AI-assisted writing process. Participant-36 ($C_{before}$) noted that *"AI has a positive impact on my thinking by generating knowledge I had never heard of before".*

*4.2.3 Autonomy and Ownership.* The Mann-Whitney U test (see Figure 5(b)) showed a significant effect of the LLM usage timing on participants' perceived autonomy ($U = 267.500, p = 0.007, r = -0.406$). Participants who wrote on their own first had significantly higher perceived autonomy in the ideation process ($C_{after}$: $M = 4.400; SD = 1.505$) than those who immediately generated ideas with LLM ($C_{before}$: $M = 3.417; SD = 1.274$).

When it comes to perceived ownership of the ideation outcomes (see Figure 5(c)), the Mann-Whitney U test also indicated a significant effect of the LLM usage timing on perceived ownership of the ideation outcome ($U = 227.000, p = 0.001, r = -0.496$).

---
[10]https://jasp-stats.org/

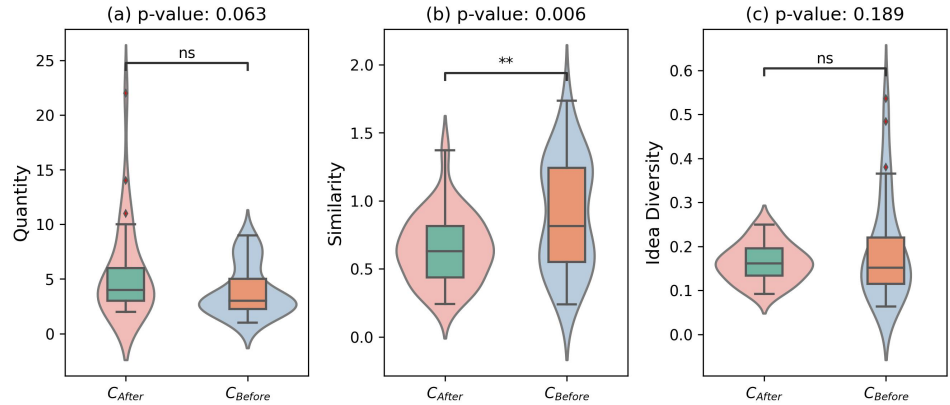

Figure 3: Comparison between $C_{before}$ and $C_{after}$ in (a) idea quantity, (b) idea similarity with LLM, and (c) participant's idea diversity. The violin plots represent the distribution, while the overlaid boxplots show the median and interquartile ranges. Significant differences between the conditions are indicated by p-value annotations, where ns >= 0.05, * < 0.05, ** < 0.01, *** < 0.001.

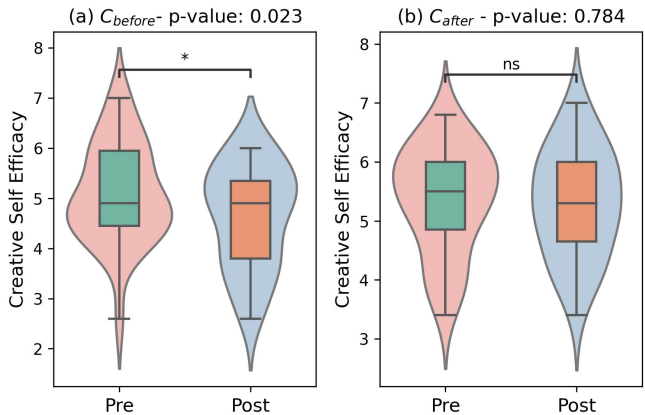

Figure 4: Comparison between pre- and post creative self-efficacy in (a) $C_{before}$ and (b) $C_{after}$ groups. The violin plots represent the distribution, while the overlaid boxplots show the median and interquartile ranges. Significant differences between the conditions are indicated by p-value annotations, where ns >= 0.05, * < 0.05, ** < 0.01, *** < 0.001. This figure shows that when LLM intervenes from the beginning $C_{before}$, there is a significant decrease in creative self-efficacy after task completion. Conversely, in $C_{after}$, the creative self-efficacy remains consistent during the creation process.

Generating ideas with LLM at the beginning decreased participants' perceived ownership ($C_{before}$: $M = 3.967; SD = 1.216$) compared with those in $C_{after}$ who used LLM after individual ideation ($M = 4.993; SD = 1.222$).

*4.2.4 Credit Attribution.* We also examined how participants assign credit to themselves and LLM when using LLM at different timings when ideating with LLM. Regarding assigning credit to participants themselves (see Figure 5(d)), the results of the Mann-Whitney U test indicated that there was a significant effect of the different timings of LLM usage ($U = 304.000, p = 0.029, r = -0.324$) on participants' attribution of credit for themselves for the ideation outcome. Participants who ideated immediately with LLM ($C_{before}$: $M = 3.367; SD = 1.608$) assigned less credit to themselves compared with those who ideated independently first and then used LLM later ($C_{after}$: $M = 4.267; SD = 1.552$).

Regarding assigning credit to LLM (see Figure 5(e)), the Mann-Whitney U test showed a significant effect of the different timings of LLM usage on participants' attribution of credit for LLM for the ideation outcome ($U = 628.500, p = 0.007, r = 0.397$). Participants who generated ideas immediately with LLM attributed more credit to LLM ($C_{before}$: $M = 5.367; SD = 1.159$) compared to those who generated ideas independently first and then used LLM later ($C_{after}$: $M = 4.233; SD = 1.675$).

*4.2.5 Different Timings of LLM Usage Influenced Autonomy, Ownership and Credit Attribution.* The qualitative findings further supported the high level of autonomy and ownership over the ideation outcomes when utilizing LLM at a later stage, as quantitative results showed. In particular, participants who generated ideas with LLM after their independent ideation (n = 9, $C_{after}$) felt a sense of autonomy in the ideation process. Furthermore, LLM was considered an auxiliary role to refine their existing content. For instance, Participant-2 ($C_{after}$) asserted that despite the advanced capabilities of the LLM, he maintained a leading role because he had already formed his viewpoints: *"I believe AI has more information than I do, so it can expand on my ideas and explain them more eloquently, but I have creative ideas as well, and in this exercise mostly, the ideas originated from me, AI [just] added to my core ideas or helped to explain them properly."* Another participant shared a similar attitude and claimed the LLM contributed a small amount of work, as Participant-13 ($C_{after}$) mentioned: *"[...] most of the written ideas were mine, I only used a few [ideas] directly from AI. The AI ideas that were used were elaborations on ideas that I had already listed."* These comments suggested that participants had high-level autonomy and tended to claim more credit for the ideation outcomes.

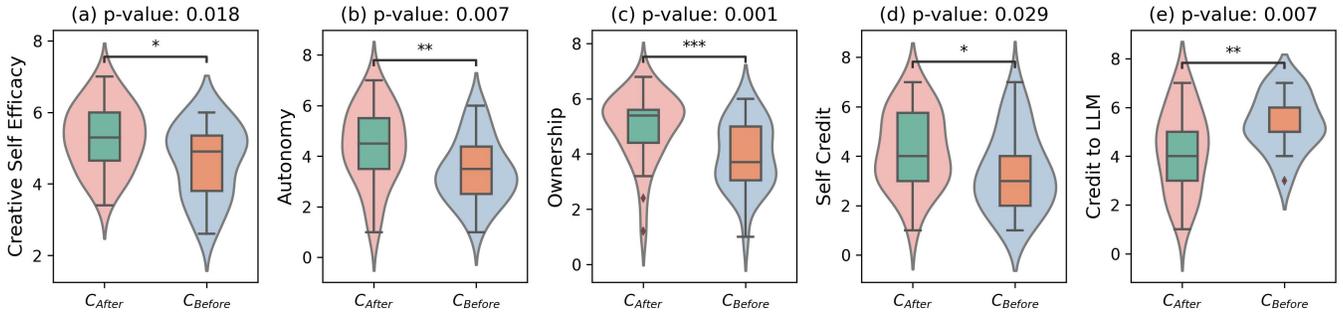

Figure 5: Comparison between $C_{before}$ and $C_{after}$ in perceptions of (a) creative self-efficacy, (b) autonomy, (c) ownership, (d) self credit, and (e) credit to LLM. The violin plots represent the distribution, while the overlaid boxplots show the median and interquartile ranges. Significant differences between the conditions are indicated by p-value annotations, where ns >= 0.05, * < 0.05, ** < 0.01, *** < 0.001.

In contrast, when the LLM was used at the beginning, many participants (n = 7, $C_{before}$ condition) reported they had a decreased sense of autonomy when generating supporting ideas. Generating ideas immediately with LLM's assistance made participants think the whole ideation task ended up being like prompt engineering. For instance, Participant-32 described, *"Well, I was just typing a prompt, and then the AI gave me the results I wanted, or if I didn't like it, I just rewrote the prompt and saw the new result" (Participant-32, $C_{before}$)*. Participant-59 ($C_{before}$) claimed that *"I had to just write a sentence which prompted really good arguments by the AI. I didn't feel like I was writing the essay at all. Simply give AI the topic to write about."* Moreover, when participants were initially presented with ideas from LLM, they struggled to add meaningful contributions, further diminishing their sense of autonomy. As Participant-58 noted, *"Most of the ideas the AI generated were excellent and were similar to ideas I was already thinking of, so the AI more or less just helped me complete my ideas to be typed in in a more coherent way, so it was difficult to think of something more creative than AI."* Participant-50 ($C_{before}$) added, *"The AI did all the writing. I just read it to make sure that it makes sense."*

## 4.3 Mediators that Influence the Timing of LLM Usage in Idea Generation and Credit Attribution (RQ3)

Regarding RQ1 and RQ2, we found that the timing of using LLM had a significant impact on both ideation outcomes and participants' perceptions. In this part, we moved further to model these factors according to previous studies (listed below) to explore how different timing of LLM usage affected participants' ideation results and credit attribution through its influence on participants' perceptions.

*4.3.1 Existing Relationships between Variables.* Research on human-AI collaboration has indicated that AI intervention can alter autonomy perception [2, 45]. Therefore, we hypothesized that using LLM at different timings could influence people's perceived autonomy. Furthermore, some studies claimed that the sense of ownership is shaped by the feeling of autonomy [18, 55]. As Bandura and Adams's self-efficacy theory suggested, an individual's belief in their ability to execute tasks influenced their motivation and performance [3, 5]. Creative self-efficacy, which refers to an individual's belief in their ability to generate creative ideas, has been proven to correlate with creative performance in various contexts [36, 58, 59]. Taken together, creative self-efficacy might predict the ideation outcome, including the quantity, as well as the similarity between participants' ideas and LLM-generated ideas[11].

Moreover, autonomy and ownership are both considered to affect creative self-efficacy, with autonomy enhancing the individual's sense of control over the creative process, and ownership fostering personal involvement and accountability [36, 42, 65]. Both autonomy and ownership are also known to influence the perception of responsibility, aligning with the idea that individuals feel more accountable for outcomes when they have a stronger sense of control and ownership [43, 52].

*4.3.2 Structural Equation Model Analysis.* To answer this RQ3 and explore the hypotheses mentioned above, we built a Structural Equation Model (SEM) using bootstrap resampling with 1000 samples to estimate the 95% confidence intervals. The model examined how LLM usage timing predicted autonomy and ownership, which in turn influenced the ideation outcomes and credit attribution. We listed all relationships modeled in SEM in Table 2 and highlighted significant results in bold.

The standardized root mean square residual (SRMR) was 0.003, indicating an excellent model fit, suggesting that the model adequately explains the data and supports the hypothesized relationships: $\chi^2(2) = 0.120, p = 0.942, CFI = 1.000, TLI = 1.126$, and $RMSEA = 0.000\ (95\%\ CI = [0.000, 0.051], p = 0.950)$. The detailed regression coefficients are shown in Table 2, and significant pathways of effects are shown in Figure 6.

The result indicated that the LLM usage timing significantly predicted the autonomy ($\beta = 0.983, SE = 0.354, p = 0.005, 95\%\ CI = [0.186, 1.636]$). When generating ideas independently before using LLM, participants' autonomy increased. Furthermore, autonomy significantly predicted the sense of ownership ($\beta = 0.632, SE = 0.078, p < 0.001, 95\%\ CI = [0.437, 0.817]$) and fully mediated the effect of LLM usage timing on ownership, showing that the improvement in participant's ownership was completely derived from their increased autonomy.

---
[11]The participant's idea diversity is excluded from the modeling analysis due to the non-significant result.

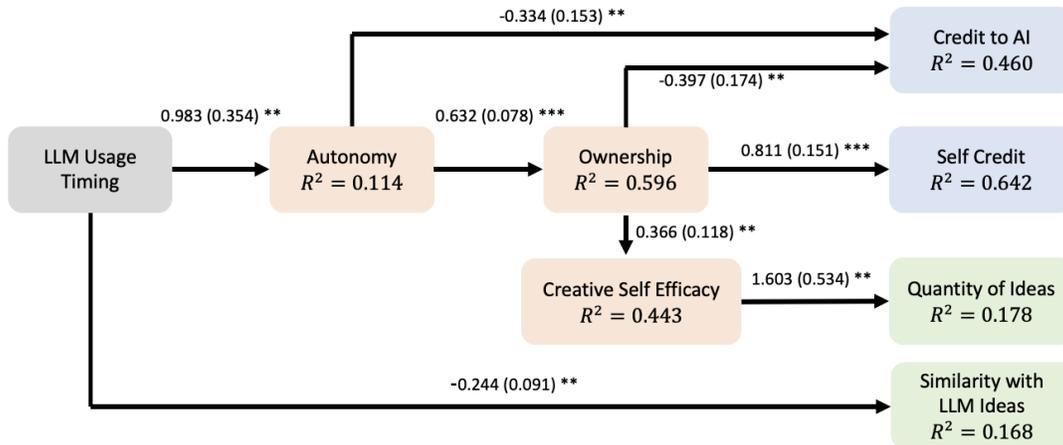

**Figure 6:** This figure illustrates the structural equation model (SEM) used to understand how LLM usage timing affects (in gray) ideation outcomes (in green) and credit attribution (in blue), where only the paths with significant effect are shown (*p < 0.05, **p < 0.01, ***p < 0.001). Please refer to Table 2 for all paths incorporated in SEM. The numbers on the arrows represent the regression coefficients, and the values in parentheses indicate the standard error (SE) associated with each estimate. Based on previous literature, this model examines autonomy, ownership, and creative self-efficacy as serial mediators (in orange), showing the direct and indirect effects of LLM usage timing on other perceptions, as well as ideation outcomes. Specifically, the LLM usage after independent ideation improves autonomy, followed by increasing the sense of ownership, which subsequently boosts creative self-efficacy and ultimately impacts ideation outcomes and credit attribution.

Additionally, ownership significantly predicted creative self-efficacy ($\beta = 0.366, SE = 0.118, p = 0.002, 95\% CI = [0.127, 0.632]$) and fully mediated the relationship between autonomy and creative self-efficacy, which indicated that the improvement in creative self-efficacy is largely due to the increase in the sense of ownership instead of autonomy's direct effect.

In parallel, ownership was a significant predictor of self-credit ($\beta = 0.811, SE = 0.151, p < 0.001, 95\% CI = [0.507, 1.135]$) and credit to LLM ($\beta = -0.397, SE = 0.174, p = 0.023, 95\% CI = [-0.708, -0.075]$). The difference was that self-credit was almost fully mediated by ownership, while credit to LLM was both affected directly by autonomy and mediated by ownership, which means that the enhancement in autonomy would (1) directly decrease the credit to LLM, and (2) decrease that by feeling more control over the results.

Finally, creative self-efficacy significantly predicted the quantity of supporting ideas generated by participants ($\beta = 1.603, SE = 0.534, p = 0.003, 95\% CI = [0.696, 2.979]$). The higher the creative self-efficacy was, the more supporting ideas participants generated. However, it cannot effectively predict the similarity with ideas from LLM ($\beta = 0.084, SE = 0.055, p = 0.125, 95\% CI = [-0.032, 0.182]$), which was solely influenced by the LLM usage timing itself ($\beta = -0.244, SE = 0.091, p = 0.007, 95\% CI = [-0.397, -0.064]$).

These findings suggested that autonomy plays a critical role in mediating the effects of LLM usage timing on ideation outcomes (quantity of supporting ideas and similarity with ideas from LLM) and credit attribution (assigning credit to oneself and the LLM). Increased autonomy, resulting from independent ideation before using LLM, enhanced participants' sense of ownership, which in turn led to an increase in self-credit and a reduction in credit-to-AI. Enhanced autonomy also increased the quantity of ideas through ownership enhancement, highlighting the importance of autonomy in maintaining the motivation for ideation. However, the similarity with ideas from LLM appeared to be directly influenced by the LLM usage timing, independent of ownership, showing that when LLM is involved from the start, the similarity between users' ideas and LLM's ideas increased.

## 5 Discussion

Our results showed that introducing LLM after participants' independent ideation led to increased autonomy, greater ownership, and higher creative self-efficacy compared to those who used LLM at the beginning of the task (see Table 3).

### 5.1 Delay LLM usage Timing to Avoid Overreliance and Protect Divergent Thinking

We found that participants who used LLM before ideating their own supporting ideas independently had higher idea similarity with LLM than those who used LLM after independent ideation. They also had significantly lower levels of perceived autonomy, ownership, self-credit, and creative self-efficacy in their ideation process compared to those who used LLM after ideating independently. Drawing from qualitative insights, we found that early introduction of LLMs may reduce the cognitive effort required for idea generation, leading participants to rely more on AI-generated content rather than engaging in deeper, independent thought. This partially aligns with previous research, which suggested that AI involvement can reduce cognitive engagement and foster overreliance on technological assistance [13, 19, 50]. However, we offer a more nuanced view, arguing that these negative consequences primarily arise when LLM support is introduced before participants have formed their own initial ideas. Once people have formulated their own ideas, LLM usage enables them to keep their original ideas without being influenced by LLM-generated ideas.

Table 2: Regression coefficients of the relationships between LLM usage timing, autonomy, ownership, credit attribution and ideation outcomes. The table also includes the standard errors (SE), z-values, p-values, and 95% confidence intervals for each predictor's estimate, highlighting statistically significant effects (*p < 0.05, **p < 0.01, ***p < 0.001, and † for marginally significant).

| Outcome | Predictor | Estimate | Std. Error | z-value | p-value | 95% CI Lower | 95% CI Upper |
|---|---|---|---|---|---|---|---|
| **Autonomy** | **LLM Usage Timing** | 0.983 | 0.354 | 2.778 | **0.005**** | 0.186 | 1.636 |
| **Ownership** | **Autonomy** | 0.632 | 0.078 | 8.087 | **<0.001**** | 0.437 | 0.817 |
|  | LLM Usage Timing | 0.406 | 0.227 | 1.784 | 0.074 | −0.121 | 0.928 |
| **Creative Self Efficacy** | **Ownership** | 0.366 | 0.118 | 3.114 | **0.002**** | 0.127 | 0.632 |
|  | Autonomy | 0.132 | 0.103 | 1.287 | 0.198 | −0.041 | 0.315 |
|  | LLM Usage Timing | 0.168 | 0.212 | 0.788 | 0.430 | −0.241 | 0.628 |
| **Quantity of Ideas** | **Creative Self Efficacy** | 1.603 | 0.534 | 3.000 | **0.003**** | 0.696 | 2.979 |
|  | LLM Usage Timing | 1.563 | 0.905 | 1.727 | 0.084 | 0.120 | 3.210 |
|  | Autonomy | −0.302 | 0.441 | −0.685 | 0.494 | −0.959 | 0.582 |
|  | Ownership | −0.629 | 0.535 | −1.175 | 0.240 | −1.527 | 0.509 |
| **Similarity with Ideas from LLM** | Creative Self Efficacy | 0.084 | 0.055 | 1.534 | 0.125 | −0.032 | 0.182 |
|  | **LLM Usage Timing** | −0.244 | 0.091 | −2.678 | **0.007**** | −0.397 | −0.064 |
|  | Autonomy | −0.042 | 0.044 | −0.942 | 0.346 | −0.130 | 0.042 |
|  | Ownership | −0.021 | 0.054 | −0.397 | 0.691 | −0.126 | 0.083 |
| **Self Credit** | **Autonomy** | 0.226 | 0.132 | 1.712 | **0.087**† | −0.056 | 0.543 |
|  | **Ownership** | 0.811 | 0.151 | 5.382 | **<0.001**** | 0.507 | 1.135 |
|  | LLM Usage Timing | −0.155 | 0.273 | −0.569 | 0.570 | −0.743 | 0.440 |
| **Credit to AI** | **Autonomy** | −0.334 | 0.153 | −2.186 | **0.029**** | −0.639 | −0.030 |
|  | **Ownership** | −0.397 | 0.174 | −2.276 | **0.023**** | −0.708 | −0.075 |
|  | LLM Usage Timing | −0.398 | 0.315 | −1.261 | 0.207 | −1.026 | 0.247 |

Table 3: Summary table of research questions and key findings.

| Research Question | Key Findings |
|---|---|
| **RQ1:** How does ideating with LLM at different timings (*before* or *after* independent ideation) influence ideation outcomes (e.g., quantity of supporting ideas, similarity with LLM ideas, and personal idea diversity)? | - Participants who used LLM after ideating independently ($C_{after}$) produced slightly more ideas than those who used LLM at the beginning ($C_{before}$), though the effect was marginally significant ($p = 0.063$). <br> - Generating ideas with LLM from the start resulted in lower originality reflected by the higher idea similarity with LLM-generated content ($p = 0.006$). |
| **RQ2:** How does ideating with LLM at different timings (*before* or *after* independent ideation) impact participants' perceptions of their creative self-efficacy, autonomy, ownership, and credit attribution? | - LLM usage timing significantly impacted creative self-efficacy: Participants in the $C_{after}$ condition demonstrated higher creative self-efficacy ($p = 0.018$). <br> - $C_{after}$ condition led to significantly higher perceived autonomy ($p = 0.007$) and ownership ($p = 0.001$) compared to $C_{before}$. <br> - Participants in the $C_{before}$ condition attributed more credit to AI, while $C_{after}$ participants gave more credit to themselves ($p = 0.029$ for self-credit, $p = 0.007$ for AI credit). |
| **RQ3:** What is the mediation mechanism, and how does it transmit the effect of LLM usage timing to ideation outcomes and credit attribution? | - Autonomy fully mediated the effect of LLM usage timing on ownership ($p < 0.001$). <br> - Ownership fully mediated the effect of autonomy on creative self-efficacy and self-credit. <br> - Creative self-efficacy mediated the relationship between ownership and idea quantity but had no significant effect on the similarity with ideas from LLM. |

We also believe that this timing effect of LLM usage can be better understood through the lens of divergent and convergent thinking in the creativity process [21, 22]. Divergent thinking is a cognitive process aimed at generating multiple innovative ideas by exploring numerous possible solutions. After this phase, convergent thinking organizes and structures these ideas logically to reach a single solution [21, 22]. When LLMs intervene too early in the ideation process, LLMs may introduce fixed logical paths, constraining people's thinking space and hindering the promotion of divergent thinking, similar to the idea fixation process [27, 35, 46, 60]. However, when LLMs are introduced *after* people have independently generated ideas, they have already developed scattered elements that can lead to many potential solutions. Therefore, LLMs may not inhibit the divergent thinking process significantly. Specifically, after divergence, participants may struggle with constructing a single conclusion due to the multitude of options. Here, LLMs can be more helpful by leveraging their ability to structure logic and filter information [63, 66], assisting creators in organizing their ideas, thus benefiting their creation outcomes, as evidenced by Participant-29's expression in Section 4.2.2.

Based on these findings, we recommend that designers of human-AI collaboration systems: (1) Delay access to LLM assistance until users have engaged in independent idea generation. For instance, LLMs could prompt users to outline their thoughts or draft initial ideas before providing AI-generated content. This would help preserve the efficiency and idea expansion benefits of LLMs while reducing the risk of overreliance. (2) Encourage LLMs to offer fragmentary ideas or elements rather than complete solutions during the initial creative phases to promote more active divergent thinking, with LLMs later aiding in the convergent phase by helping users organize and evaluate their ideas, ultimately fostering creativity.

## 5.2 Maintaining User Autonomy and Idea Ownership when Collaborating with LLM

The timing of the LLM usage altered participants' perception of autonomy, subsequently impacting their sense of ownership, creative self-efficacy, credit attribution, and ideation outcomes. This phenomenon aligns with the assertions made by Deci and Ryan in self-determination theory [12], underscoring that autonomy is crucial to fostering intrinsic motivation and improving individual self-efficacy. Additionally, in McGuire et al.'s study [38], they found that when the writer acts as an editor of the LLM, it diminishes their writing outcomes. However, when the writer and LLM act as collaborators, this negative effect is mitigated. They claimed that *"people must occupy the role of a co-creator, not an editor, to reap the benefits of generative AI in the production of creative works"* and speculated the results might be interpreted by autonomy perception. Our findings extend this research in two ways: (1) we empirically demonstrate that autonomy and ownership are key mediating factors fully driving creative self-efficacy, ultimately shaping the creative outcome; (2) Our findings expand McGuire et al.'s claim, suggesting that as long as the collaboration between human and LLM protects user autonomy, the creative outcome can be safeguarded instead of *"must be co-creator"*.

*5.2.1 Introducing an Individual Idea Incubation Mode in Human-AI Collaboration.* When developing human-AI collaboration systems, a phased approach [29, 40] can be employed where the level of AI involvement is carefully controlled at different stages for the given task. For instance, during the initial phase, AI can pause temporarily or encourage users to think alone first without actively generating output for them. As the creative process progresses, the AI's involvement can become more active, encouraging users to refine and expand upon their own ideas. This approach could help prevent idea fixation due to early over-reliance [17, 56] on AI while maintaining the user's sense of ownership.

*5.2.2 Adapt to Users' Perceived Autonomy Dynamically.* As our result showed that autonomy was the vital mediator for credit attribution and ideation outcomes (Section 4.3), future AI systems could introduce a dynamic autonomy adjustment mechanism that allows the AI's involvement to be tailored in real-time based on the user's perceived autonomy when performing a task. This dynamic adjustment ensures that users retain a balanced level of autonomy throughout the individual ideation process, leading to enhanced ownership of their work. This is especially important as different levels of ownership perceived by users will further influence how much credit they assign to AI and themselves when using AI to complete a given task.

## 6 Limitation and future works

Though we identified the importance of the timing of LLM usage in AI-assisted ideation, we acknowledge several limitations in our study. First, the specific writing tasks, a health-related topic, and the LLM used in this study could have influenced the outcomes. While our task required participants to generate supporting ideas based on a given topic, relying more on their existing knowledge and logical reasoning than on a deep understanding of specialized subject matter, we cannot rule out the possibility that differences in people's background knowledge may have impacted their ideation outcomes. Controlling people's prior understanding related to the ideation task in future studies could enhance our overall understanding.

Moreover, to enable the feasibility of an online controlled experiment, we designed the ideation task in a relatively simple way, in which we invited participants to generate supporting ideas for the given topic. However, in more complex tasks that demand higher levels of creativity and critical thinking, the timing of LLM usage may have different effects. This influence may also vary across different types of creative tasks, such as image creation, design, and other AI-assisted creative activities. Future research could explore how the timing of LLM usage affects the creative process and outcomes in tasks of varying complexity and across diverse domains.

Finally, while we employed a controlled experimental design, several confounding variables—such as participants' prior experience with AI, individual differences in writing skills, and familiarity with the writing tasks—may play different roles in the AI-assisted writing process. Future studies are encouraged to investigate the potential influence of these factors further.

# 7 Conclusion

This research explores the impact of using LLM at different timings on individuals' perceptions and ideation outcomes, using an ideation task. The findings revealed that introducing LLMs after independent ideation led people to have lowered similarity between their ideas and LLM-generated ideas. It also fostered people's sense of autonomy, ownership, and creative self-efficacy during the ideation process. The analysis of SEM demonstrated that increasing autonomy empowered people with a sense of control over their ideation outcomes, leading to increased creative outputs. Conversely, using LLM before independent ideation may lead to over-reliance on AI-generated content, suppressing divergent thinking and even undermining individuals' creative self-efficacy. We argue that incorporating an individual idea incubation stage in AI-assisted ideation could enhance the impact of LLM in supporting idea expansion without succumbing to idea fixation by LLM.

## Acknowledgments

This research work is partially supported by NUS IT's Research Computing group (NUSREC-HPC-00001) and AWS.

## References

[1] Barrett R Anderson, Jash Hemant Shah, and Max Kreminski. 2024. Homogenization effects of large language models on human creative ideation. In *Proceedings of the 16th conference on creativity & cognition*. 413–425.
[2] Quentin André, Ziv Carmon, Klaus Wertenbroch, Alia Crum, Douglas Frank, William Goldstein, Joel Huber, Leaf Van Boven, Bernd Weber, and Haiyang Yang. 2018. Consumer choice and autonomy in the age of artificial intelligence and big data. *Customer needs and solutions* 5 (2018), 28–37.
[3] A Bandura. 1997. Self-efficacy: The Exercise of Control. New York: W. H Freeman. *Self-efficacy: the exercise of control. New York: WH Freeman* (1997).
[4] Albert Bandura et al. 1986. Social foundations of thought and action. *Englewood Cliffs, NJ* 1986, 23-28 (1986), 2.
[5] Albert Bandura and Nancy E Adams. 1977. Analysis of self-efficacy theory of behavioral change. *Cognitive therapy and research* 1, 4 (1977), 287–310.
[6] Virginia Braun and Victoria Clarke. 2006. Using thematic analysis in psychology. *Qualitative research in psychology* 3, 2 (2006), 77–101.
[7] Lijia Chen, Pingping Chen, and Zhijian Lin. 2020. Artificial intelligence in education: A review. *Ieee Access* 8 (2020), 75264–75278.
[8] Li-Yuan Chiou, Peng-Kai Hung, Rung-Huei Liang, and Chun-Teng Wang. 2023. Designing with AI: an exploration of co-ideation with image generators. In *Proceedings of the 2023 ACM designing interactive systems conference*. 1941–1954.
[9] Elizabeth Clark, Anne Spencer Ross, Chenhao Tan, Yangfeng Ji, and Noah A Smith. 2018. Creative writing with a machine in the loop: Case studies on slogans and stories. In *23rd International Conference on Intelligent User Interfaces*. 329–340.
[10] Mark Coeckelbergh. 2020. Artificial intelligence, responsibility attribution, and a relational justification of explainability. *Science and engineering ethics* 26, 4 (2020), 2051–2068.
[11] Samuel Rhys Cox, Yunlong Wang, Ashraf Abdul, Christian Von Der Weth, and Brian Y. Lim. 2021. Directed diversity: Leveraging language embedding distances for collective creativity in crowd ideation. In *Proceedings of the 2021 CHI Conference on Human Factors in Computing Systems*. 1–35.
[12] Edward L Deci and Richard M Ryan. 2012. Self-determination theory. *Handbook of theories of social psychology* 1, 20 (2012), 416–436.
[13] Ismail Dergaa, Helmi Ben Saad, Jordan M Glenn, Badii Amamou, Mohamed Ben Aissa, Noomen Guelmami, Feten Fekih-Romdhane, and Karim Chamari. 2024. From tools to threats: a reflection on the impact of artificial-intelligence chatbots on cognitive health. *Frontiers in Psychology* 15 (2024), 1259845.
[14] Zijian Ding, Arvind Srinivasan, Stephen MacNeil, and Joel Chan. 2023. Fluid transformers and creative analogies: Exploring large language models' capacity for augmenting cross-domain analogical creativity. In *Proceedings of the 15th Conference on Creativity and Cognition*. 489–505.
[15] Anil R Doshi and Oliver Hauser. 2023. Generative artificial intelligence enhances creativity. *Available at SSRN* (2023).
[16] Fiona Draxler, Anna Werner, Florian Lehmann, Matthias Hoppe, Albrecht Schmidt, Daniel Buschek, and Robin Welsch. 2023. The AI Ghostwriter Effect: Users Do Not Perceive Ownership of AI-Generated Text But Self-Declare as Authors. *arXiv preprint arXiv:2303.03283* (2023).

[17] Andrew R du Rocher. 2020. Active learning strategies and academic self-efficacy relate to both attentional control and attitudes towards plagiarism. *Active Learning in Higher Education* 21, 3 (2020), 203–216.
[18] Marylène Gagné and Edward L Deci. 2005. Self-determination theory and work motivation. *Journal of Organizational behavior* 26, 4 (2005), 331–362.
[19] A Shaji George, T Baskar, and P Balaji Srikaanth. 2024. The Erosion of Cognitive Skills in the Technological Age: How Reliance on Technology Impacts Critical Thinking, Problem-Solving, and Creativity. (2024).
[20] Katy Ilonka Gero and Lydia B Chilton. 2019. Metaphoria: An algorithmic companion for metaphor creation. In *Proceedings of the 2019 CHI conference on human factors in computing systems*. 1–12.
[21] J Guilford. 1950. Creativity. American Psychology. 5 (9), 444–454.
[22] Joy Paul Guilford. 1967. The nature of human intelligence. *New York: Macgraw Hill* (1967).
[23] Shirley Hayati, Minhwa Lee, Dheeraj Rajagopal, and Dongyeop Kang. 2024. How Far Can We Extract Diverse Perspectives from Large Language Models?. In *Proceedings of the 2024 Conference on Empirical Methods in Natural Language Processing*. 5336–5366.
[24] Cheng-Zhi Anna Huang, Hendrik Vincent Koops, Ed Newton-Rex, Monica Dinculescu, and Carrie J Cai. 2020. AI song contest: Human-AI co-creation in songwriting. *arXiv preprint arXiv:2010.05388* (2020).
[25] Angel Hsing-Chi Hwang. 2022. Too late to be creative? AI-empowered tools in creative processes. In *CHI Conference on Human Factors in Computing Systems Extended Abstracts*. 1–9.
[26] Maurice Jakesch, Advait Bhat, Daniel Buschek, Lior Zalmanson, and Mor Naaman. 2023. Co-writing with opinionated language models affects users' views. In *Proceedings of the 2023 CHI Conference on Human Factors in Computing Systems*. 1–15.
[27] David G Jansson and Steven M Smith. 1991. Design fixation. *Design studies* 12, 1 (1991), 3–11.
[28] Kimberly S Jaussi, Amy E Randel, and Shelley D Dionne. 2007. I am, I think I can, and I do: The role of personal identity, self-efficacy, and cross-application of experiences in creativity at work. *Creativity Research Journal* 19, 2-3 (2007), 247–258.
[29] Kenneth R Koedinger, John R Anderson, William H Hadley, and Mary A Mark. 1997. Intelligent tutoring goes to school in the big city. *International Journal of Artificial Intelligence in Education* 8 (1997), 30–43.
[30] Nicholas W Kohn and Steven M Smith. 2011. Collaborative fixation: Effects of others' ideas on brainstorming. *Applied Cognitive Psychology* 25, 3 (2011), 359–371.
[31] Chinmay Kulkarni, Steven P Dow, and Scott R Klemmer. 2013. Early and repeated exposure to examples improves creative work. In *Design thinking research: Building innovation eco-systems*. Springer, 49–62.
[32] Arto Laitinen and Otto Sahlgren. 2021. AI systems and respect for human autonomy. *Frontiers in artificial intelligence* 4 (2021), 705164.
[33] Bart Lamiroy and Emmanuelle Potier. 2022. Lamuse: Leveraging Artificial Intelligence for Sparking Inspiration. In *International Conference on Computational Intelligence in Music, Sound, Art and Design (Part of EvoStar)*. Springer, 148–161.
[34] Mina Lee, Percy Liang, and Qian Yang. 2022. Coauthor: Designing a human-ai collaborative writing dataset for exploring language model capabilities. In *Proceedings of the 2022 CHI conference on human factors in computing systems*. 1–19.
[35] Julie S Linsey, Ian Tseng, Katherine Fu, Jonathan Cagan, Kristin L Wood, and Christian Schunn. 2010. A study of design fixation, its mitigation and perception in engineering design faculty. (2010).
[36] Gro Ellen Mathisen. 2011. Organizational antecedents of creative self-efficacy. *Creativity and Innovation Management* 20, 3 (2011), 185–195.
[37] Marian Mazzone and Ahmed Elgammal. 2019. Art, creativity, and the potential of artificial intelligence. In *Arts*, Vol. 8. MDPI, 26.
[38] Jack McGuire, David De Cremer, and Tim Van de Cruys. 2024. Establishing the importance of co-creation and self-efficacy in creative collaboration with artificial intelligence. *Scientific Reports* 14, 1 (2024), 18525.
[39] Piotr Mirowski, Kory W Mathewson, Jaylen Pittman, and Richard Evans. 2023. Co-Writing Screenplays and Theatre Scripts with Language Models: Evaluation by Industry Professionals. In *Proceedings of the 2023 CHI Conference on Human Factors in Computing Systems*. 1–34.
[40] Antonija Mitrovic, Brent Martin, and Michael Mayo. 2002. Using evaluation to shape ITS design: Results and experiences with SQL-Tutor. *User Modeling and User-Adapted Interaction* 12 (2002), 243–279.
[41] Bernard A Nijstad and Wolfgang Stroebe. 2006. How the group affects the mind: A cognitive model of idea generation in groups. *Personality and social psychology review* 10, 3 (2006), 186–213.
[42] Chiyem Lucky Nwanzu and Sunday Samson Babalola. 2022. The influence of psychological ownership and creative self-efficacy on employee creative performance. *Pakistan Journal of Commerce and Social Sciences (PJCSS)* 16, 1 (2022), 34–49.
[43] Jon L Pierce, Tatiana Kostova, and Kurt T Dirks. 2001. Toward a theory of psychological ownership in organizations. *Academy of management review* 26, 2

(2001), 298–310.
[44] Reza Pourreza, Apratim Bhattacharyya, Sunny Panchal, Mingu Lee, Pulkit Madan, and Roland Memisevic. 2023. Painter: Teaching Auto-regressive Language Models to Draw Sketches. *arXiv preprint arXiv:2308.08520* (2023).
[45] Carina Prunkl. 2024. Human Autonomy at Risk? An Analysis of the Challenges from AI. *Minds and Machines* 34, 3 (2024), 26.
[46] A Terry Purcell and John S Gero. 1996. Design and other types of fixation. *Design studies* 17, 4 (1996), 363–383.
[47] Othman Sbai, Mohamed Elhoseiny, Antoine Bordes, Yann LeCun, and Camille Couprie. 2018. Design: Design inspiration from generative networks. In *Proceedings of the European Conference on Computer Vision (ECCV) Workshops*. 0–0.
[48] Daniel L Schwartz, Catherine C Chase, Marily A Oppezzo, and Doris B Chin. 2011. Practicing versus inventing with contrasting cases: The effects of telling first on learning and transfer. *Journal of educational psychology* 103, 4 (2011), 759.
[49] Daniel L Schwartz and Taylor Martin. 2004. Inventing to prepare for future learning: The hidden efficiency of encouraging original student production in statistics instruction. *Cognition and instruction* 22, 2 (2004), 129–184.
[50] Mathura Shanmugasundaram and Arunkumar Tamilarasu. 2023. The impact of digital technology, social media, and artificial intelligence on cognitive functions: a review. *Frontiers in Cognition* 2 (2023), 1203077.
[51] Pao Siangliulue, Joel Chan, Krzysztof Z Gajos, and Steven P Dow. 2015. Providing timely examples improves the quantity and quality of generated ideas. In *Proceedings of the 2015 ACM SIGCHI Conference on Creativity and Cognition*. 83–92.
[52] Monika Simmler. 2024. Responsibility gap or responsibility shift? The attribution of criminal responsibility in human–machine interaction. *Information, Communication & Society* 27, 6 (2024), 1142–1162.
[53] Dilpreet Singh, Nina Rajcic, Simon Colton, and Jon McCormack. 2019. Camera obscurer: generative art for design inspiration. In *Computational Intelligence in Music, Sound, Art and Design: 8th International Conference, EvoMUSART 2019, Held as Part of EvoStar 2019, Leipzig, Germany, April 24–26, 2019, Proceedings 8*. Springer, 51–68.
[54] Nikhil Singh, Guillermo Bernal, Daria Savchenko, and Elena L Glassman. 2022. Where to hide a stolen elephant: Leaps in creative writing with multimodal machine intelligence. *ACM Transactions on Computer-Human Interaction* (2022).
[55] Christina Starmans. 2023. Autonomy, the moral circle, and the limits of ownership. *Behavioral & Brain Sciences* 46 (2023).
[56] Jonathan Stolk and Janie Harari. 2014. Student motivations as predictors of high-level cognitions in project-based classrooms. *Active Learning in Higher Education* 15, 3 (2014), 231–247.
[57] Timo Strohmann, Dominik Siemon, and Susanne Robra-Bissantz. 2017. brAInstorm: Intelligent assistance in group idea generation. In *Designing the Digital Transformation: 12th International Conference, DESRIST 2017, Karlsruhe, Germany, May 30–June 1, 2017, Proceedings 12*. Springer, 457–461.
[58] Min Tang, Weiping Hu, and Huan Zhang. 2017. Creative self-efficacy from the Chinese perspective: Review of studies in mainland China, Hong Kong, Taiwan, and Singapore. *The creative self* (2017), 237–257.
[59] Pamela Tierney and Steven M Farmer. 2002. Creative self-efficacy: Its potential antecedents and relationship to creative performance. *Academy of Management journal* 45, 6 (2002), 1137–1148.
[60] Luis A Vasconcelos and Nathan Crilly. 2016. Inspiration and fixation: Questions, methods, findings, and challenges. *Design Studies* 42 (2016), 1–32.
[61] Luis A Vasconcelos, Maria A Neroni, Carlos Cardoso, and Nathan Crilly. 2018. Idea representation and elaboration in design inspiration and fixation experiments. *International Journal of Design Creativity and Innovation* 6, 1-2 (2018), 93–113.
[62] Samangi Wadinambiarachchi, Ryan M Kelly, Saumya Pareek, Qiushi Zhou, and Eduardo Velloso. 2024. The Effects of Generative AI on Design Fixation and Divergent Thinking. *arXiv preprint arXiv:2403.11164* (2024).
[63] Jason Wei, Xuezhi Wang, Dale Schuurmans, Maarten Bosma, Fei Xia, Ed Chi, Quoc V Le, Denny Zhou, et al. 2022. Chain-of-thought prompting elicits reasoning in large language models. *Advances in neural information processing systems* 35 (2022), 24824–24837.
[64] Roosa Wingström, Johanna Hautala, and Riina Lundman. 2023. Redefining creativity in the era of AI? Perspectives of computer scientists and new media artists. *Creativity Research Journal* (2023), 1–17.
[65] CHEN YANG. 2024. Teacher Autonomy, Creative Self-efficacy, and Innovative Behavior: Perspectives from Chinese University EFL Teachers. *Arab World English Journal* 15, 2 (2024).
[66] Nicolas Yax, Hernan Anlló, and Stefano Palminteri. 2024. Studying and improving reasoning in humans and machines. *Communications Psychology* 2, 1 (2024), 51.
[67] Robert J Youmans and Thomaz Arciszewski. 2014. Design fixation: Classifications and modern methods of prevention. *AI EDAM* 28, 2 (2014), 129–137.
[68] Ann Yuan, Andy Coenen, Emily Reif, and Daphne Ippolito. 2022. Wordcraft: story writing with large language models. In *27th International Conference on Intelligent User Interfaces*. 841–852.
[69] Fan Zhuo. 2021. Human-machine co-creation on artistic paintings. In *2021 IEEE 1st International Conference on Digital Twins and Parallel Intelligence (DTPI)*. IEEE, 316–319.

## A Supplementary Materials

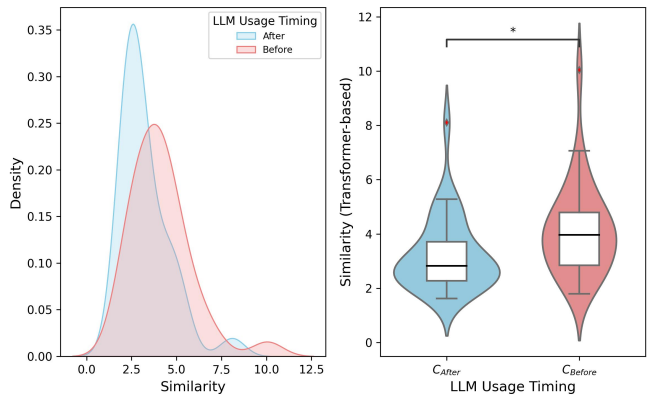

Figure 7: Comparison between $C_{before}$ and $C_{after}$ in idea similarity with LLM. (a) represents the detailed similarity distribution, where the distribution of ideas generated by participants in $C_{after}$ are concentrated in areas with lower scores while that in $C_{before}$ are comparatively higher. (b) depicts that there is a significant difference caused by LLM usage timing, which is consistent with the results in Figure 3. Significant differences between the conditions are indicated by p-value annotations, where ns >= 0.05, * < 0.05, ** < 0.01, *** < 0.001.

Table 4: An example of the ideation process using the experimental platform, which includes the prompt, participant-generated, and LLM-generated ideas on the topic of the 'jellyfish sting treatment'.

**Experimental Condition:** Condition: $C_{before}$, **Topic:** *Jellyfish Sting Treatment*
Participants were tasked to independently generate ideas and optionally use LLM for additional support.

**Prompts:** '[PROMPT] applying your urine can worse the pain for the jellyfish and it might spread the venom to other areas of the jellyfish [RESULT]'

**Participant-Generated Ideas:**
1. Applying your urine can worsen the pain for the jellyfish, and it might spread the venom to other areas of the jellyfish.
2. Urine has a slightly acidic to neutral pH, while jellyfish venom often contains proteins and toxins that are sensitive to pH changes. Applying urine to a jellyfish sting can disrupt the delicate balance of the venom's pH and potentially trigger the release of more toxins. This can intensify the pain and inflammation, making the sting worse instead of providing relief.
3. Urine may temporarily provide a cooling sensation due to its moisture, creating a placebo effect that the pain is reducing. However, it doesn't contain any specialized compounds or properties that can effectively counteract the venom's effects. Proper first aid for jellyfish stings involves rinsing the affected area with vinegar to help neutralize the toxins and using hot water to ease pain by deactivating the venom. Applying urine lacks these targeted actions and fails to address the actual components causing the pain.

**LLM-Generated Ideas:**
1. When a person applies urine to a jellyfish sting, it may spread the venom to unaffected areas, leading to a larger portion of the skin becoming affected and more painful.
2. The pH level of urine varies, making it unpredictable and unreliable for neutralizing jellyfish venom, whereas vinegar has a consistent acidity that works well.
3. Urine may also introduce bacteria to the wound, increasing the risk of infection, especially if the person has urinary tract infections or other medical conditions.
4. Despite the popular belief that urine can relieve the pain of a jellyfish sting, research suggests that it may actually worsen the situation. Urine can contain substances that further irritate the jellyfish venom and cause additional discomfort.
5. Urinating on a jellyfish sting may not provide any significant relief. The high salt content in urine can potentially increase the pain by dehydrating the affected area and intensifying the sting sensation.
6. Urine may introduce bacteria or other harmful substances to the jellyfish sting, leading to an increased risk of infection. This can result in further complications and prolong the healing process.
7. Relying on urinating as a solution for jellyfish stings may delay proper treatment. Instead of seeking medical attention or using proven remedies like vinegar or seawater to neutralize the venom, wasting time on urine application can prevent timely and effective intervention.
8. Applying urine to a jellyfish sting can actually worsen the pain and potentially spread the venom to other areas of the jellyfish. Jellyfish stings are caused by specialized cells on their tentacles called nematocysts, which inject venom into the skin upon contact.
9. Proper first aid for jellyfish stings involves rinsing the affected area with seawater, carefully removing any tentacles using tweezers or a gloved hand, and soaking the area in hot water for pain relief. Vinegar is effective in neutralizing toxins, while urine lacks such properties.